\documentclass[a4paper,12pt]{article}
\usepackage{cite}
\usepackage{amssymb,amsmath}
\newcommand{\be}{\begin{equation}}
\newcommand{\ee}{\end{equation}}
\newcommand{\bea}{\begin{eqnarray}}

\newcommand{\eea}{\end{eqnarray}}

\begin{document}

\title{On the Kerr Quantum Area Spectrum}

\author{A.J.M. Medved \\ \\
Physics Department \\
University of Seoul \\
 Seoul 130-743\\
Korea \\
E-Mail(1): allan@physics.uos.ac.kr \\
E-Mail(2):  joey\_medved@yahoo.com \\ \\}

\maketitle
\begin{abstract}

Suppose that there is a quantum operator
that describes the horizon area of a black hole.
Then what would be the form of the ensuing quantum spectrum?
In this regard, it has been conjectured that
the characteristic frequencies of the  black hole oscillations
can be used to calibrate the spacing between
the spectral levels. The current article begins with 
a brief review of  this conjecture and some of its
subsequent developments.
We then  suggest a simple but vital modification
to a recent treatment on  the Kerr (or rotating black hole)
spectrum.
As a consequence of this refinement,
we are able to rectify a prior inconsistency
(as was found between two distinct calculations) and to establish,
unambiguously, a universal form for the Kerr
and Schwarzschild spectra.

\end{abstract}
\newpage
\section*{}

There is a long-standing belief that a black hole horizon should be endowed with
a quantum area spectrum. This notion can have its origins traced back
to the profound revelations of Bekenstein in the early seventies \cite{bek-1}.
Most significantly,  the horizon area ($A$) of
a black hole was shown to be an adiabatic invariant \cite{bek-2} and, as such, should
logically obtain a discrete spectrum upon a suitable process of
 quantization. With a few other
pertinent inputs, Bekenstein went on to propose his now-famous
evenly spaced
spectrum:
\be
A_n \;=\;\epsilon \hbar n\;\;\;{\rm where}\;\;\;n=0,1,2,...\;.
\label{1}
\ee
Here, $n$ is the associated quantum number, all fundamental constants
besides Planck's constant have been set to one and
$\epsilon$ should be regarded as a  numerical coefficient of the
order of unity.

[Various
schools of thought have  questioned the viability of an evenly spaced
area spectrum; most notably, many proponents of loop quantum gravity \cite{lqg}.
It should, however, be pointed out that the rest of the discussion
focuses on the semi-classical or large $n$ limit. Hence, one
may --- if so preferred --- regard Eq.(\ref{1}) as an asymptotic form of
a possibly more convoluted spectrum.]

Along with  the form of the spectrum itself, the value of  $\epsilon$
has been somewhat controversial. On this point,  Hod made an interesting
suggestion; namely, $\epsilon$ can be fixed
 by utilizing  the  {\it quasi-normal mode} frequencies
of an oscillating
 black hole \cite{hod-1}. The premise of this so-called  {\it Hod Conjecture}
goes as follows: A perturbed black hole will tend to
 equilibrate by radiating away energy in the form of gravitational
waves. This radiation will, predominantly, consist of  damped oscillations that can be
characterized by a  discrete set of  complex frequencies. Then, ``in the spirit
of the {\it Bohr Correspondence Principle}'', it becomes natural to associate the
{\it classical} limit of these frequencies (say $\omega_c$) with the {\it large
$n$} limit  of the aforementioned (quantum) area spectrum.  Inasmuch as a given
frequency  represents a transition between spectral levels,
this association can be made rather precise:
\be
A_{n+1}-A_{n}\;=\;{\partial A(M)\over \partial M}\hbar\omega_c\;=\;32\pi M\hbar\omega_c
\;\;\; {\rm as} \;\;\; n\rightarrow\infty \;,
\label{2}
\ee
where $M$ is the black hole mass and
the well-known Schwarzschild  relation $A=16\pi M^2$ has been invoked.
[We limit considerations, for the time being, to the
Schwarzschild case of  a non-rotating (and uncharged) black hole.
The rotating or Kerr scenario will be addressed in due course.]
Note that  this difference in levels is, in Planck units, quite simply
the numerical coefficient $\epsilon$.

Hod went on to propose that, from the quasi-normal-mode side of things,
 the relevant classical limit  should be that of
infinite damping (or vanishing decay time); with the logic being that  a truly classical
black hole can not conceivably emit any radiation. More
specifically, he suggested that $\omega_c$ should be regarded as the
real part of the quasi-normal frequency after the imaginary part
(which determines the damping rate)
has been sent off to infinity. Using the
numerical evidence of the time
\cite{num-1} (which was later verified  by analytic means \cite{motl}),
Hod identified the asymptotic form for the mode frequencies as
\be
\hbar\omega(k) \;\approx\; i 2\pi T k \;+ \; T\ln 3 \;\;\; {\rm as} \;\;\; k \rightarrow\infty
\;,
\label{3}
\ee
where $k$ is the quasi-normal ``discreteness index'' and $T=\hbar /8\pi M\;$ is
the Hawking (Schwarzschild) temperature \cite{haw-1}. Then,
by way of the above line of reasoning,
\be
\hbar\omega_c\;=\; T\ln 3 \;=\;{\hbar\ln 3 \over 8\pi M}\;.
\label{4}
\ee
Finally, combining  Eqs.(\ref{2}) and (\ref{4}),  one
can reproduce Hod's important realization:
\be
\epsilon \;=\; 32\pi M {\cdot}{\ln 3 \over 8\pi M} \;=\; 4\ln3 \;.
\label{5}
\ee

Kunstatter furthered this tapestry of ideas with the following insightful
observations \cite{kun-1}:
A natural adiabatic invariant for a system with energy $E$ and vibrational
frequency $\omega(E)$ is the ratio $E/\omega(E)$. So, by way of the
{\it Bohr--Sommerfeld quantization condition}
(which essentially relates an adiabatic invariant
to a quantum number $n$ in the semi-classical or large $n$ limit),
 it follows that
\be
\int{dE\over \omega(E)}  \;\approx\; n\hbar  \;\;\; {\rm as}
\;\;\; n\rightarrow\infty\;.
\label{6}
\ee
Casting  a Schwarzschild black hole into this framework, Kunstatter
replaced $E$ with $M$ and identified  $\omega_c$ as the
most appropriate choice for the frequency.
Making the recommended substitutions and then integrating, one readily obtains
\be
{4\pi M^2 \over \ln 3}\;\approx\; n\hbar \;.
\label{7}
\ee
This obviously implies the spectral form
\be
A_n\;=\; 4n\hbar\ln 3
\label{8}
\ee
or $\epsilon=4\ln 3$ in agreement with Hod.

Recently, Maggiore refined the Hod treatment by arguing
that --- insofar as a black hole is to be viewed as a damped
harmonic oscillator --- the physically relevant frequency
would actually be \cite{mag}
\be
\omega_p \;=\; \sqrt{\omega_R^2+\left|\omega_I^2\right|}\;,
\label{9}
\ee
where $\omega_R$ and $\omega_I$ are, respectively, the
real and imaginary part of the quasi-normal mode frequency.
In the large-damping  limit, this ``physical frequency'' necessarily translates
into [{\it cf}, Eq.(\ref{3})]
\be
\hbar\omega_p \;\approx\; \hbar\left|\omega_I\right|\;\approx\;
2\pi T k \;=\;
{\hbar k\over 4M}\;\;\;
{\rm as}\;\;\; k\rightarrow\infty\;.
\label{10}
\ee
To effectively regulate this large $k$ divergence, Maggiore  sensibly
proposed that the characteristic classical frequency (our $\omega_c$)
should now be identified with a {\it transition}  between
(adjacent) quasi-normal frequency   levels. That is,
\bea
\omega_c\;&=&\;\omega_p(k+1)-\omega_p(k) \;\;\; {\rm as} \;\;\;
k\rightarrow\infty
\nonumber \\
&\approx& \; {1\over 4M} \;.
\label{11}
\eea
It should not be difficult to convince oneself that --- given this revised
state of affairs ---
the spectral-spacing coefficient now becomes
\be
\epsilon \;=\; 8\pi \;.
\label{12}
\ee
This outcome can easily  be  verified by either  Hod or Kunstatter's
choice of methodology.

[It may be of interest that this particular value
of $\epsilon$ occurs frequently  in the literature for a variety
of analytic techniques (see \cite{med}
and references therein). Many other ``selling points'' for
Maggiore's revision are discussed in
his already-cited work.
Conversely, the main attribute of the original
Hod calculation is that it complies with $\epsilon=4\ln{m}$
(for some positive integer $m>1$). This being the unique
form that complies with a strict statistical interpretation
for black hole thermodynamics \cite{bek-3};
as dictated by the area--entropy law ($S=A/4\hbar$ \cite{bek-4,haw-1}) along
with
Boltzmann--Einstein statistics.
 On the other
hand, there are  still viable reasons \cite{med,mag} why this last point might
 reasonably be overlooked.]

Even more recently, Vagenas applied the above machinery to
the Kerr or rotating (but
still uncharged) black hole case
\cite{vag}.~\footnote{Other (not necessarily related)
 derivations of the Kerr
quantum area spectrum include \cite{mak,gour2,kis}.}
It is worthwhile
to first recall the relevant thermodynamic quantities
for this scenario:
\be
A\;=\; 8\pi\left(M^2+\sqrt{M^4-J^2}\right)\;,
\label{13}
\ee
\be
T\;=\; \hbar{\sqrt{M^4-J^2}\over4\pi M \left(M^2+\sqrt{M^4-J^2}\right)}\;,
\label{14}
\ee
\be
\Omega\;=\;{J\over 2M\left(M^2+\sqrt{M^4-J^2}\right)}\;.
\label{15}
\ee
Here, $J$ is the angular momentum of the black hole
and $\Omega$ is the rotational velocity at its horizon.
(Both of which will be regarded as positive, without any loss of generality.)

Also of relevance to the current discussion,  there
has been some recent progress in understanding the
highly damped quasi-normal frequencies for
 the Kerr solution.
In fact, an analytic
result has now
 been obtained \cite{hod-2} and, reassuringly, this computation
complies with some older numerical work \cite{num-2}.
Most pertinently, the imaginary part  of
the frequency ascends asymptotically as
\be
\hbar\left|\omega_{I}\right|\;\approx 2\pi T_0 k \;\;\; {\rm as} \;\;\;
k\rightarrow\infty \;,
\label{16}
\ee
whereas the real part remains finite.
The parameter $T_0$ can be viewed as an  effective ``temperature''
and is defined by $T_0\equiv T(J=0)=\hbar /8\pi M$.

Because of the simplicity of Eq.(\ref{16}), it is quite evident
that the Kerr analogue of  Maggiore's revised (Hod) calculation
goes through unfettered;
so that, once again,
\be
\epsilon \;=\; 8\pi \;.
\label{17}
\ee
Indeed, it is this very universality that was the main observation
made by Vagenas. Unfortunately, the same type of consistency
failed to transpire for the Kerr analogue of the Kunstatter
calculation.

Given the serious nature of the last claim,
let us elaborate much further on this issue. With cognizance
of the first law of black hole thermodynamics \cite{haw-2},
Vagenas perceptively deduced the Kerr analogue
of Kunstatter's adiabatic-invariant integral as  being
\be
\int{dM-\Omega dJ\over \omega_c} \;.
\label{18}
\ee
Utilizing Eqs.(\ref{15},\ref{16}) and the Bohr--Sommerfeld quantization
condition, one promptly obtains
\be
2\int\left[2MdM \;-\;{JdJ\over M^2+\sqrt{M^4-J^2}}\right]\;\approx \;
n\hbar \;\;\; {\rm as} \;\;\;
n\rightarrow\infty \;.
\label{19}
\ee
Some straightforward integration  then reveals 
that~\footnote{We obtain an $\hbar$ in the denominator
of the logarithm  by 
cutting the integration off at the Planck scale; this being
a very  natural choice given our ignorance of sub-Planckian 
physics.  Nothing that is
said below will, however,  depend on the precise value of
this ultraviolet regulator.}
\be
2M^2 +2\sqrt{M^4-J^2}\;-\;2M^2\ln\left[{M^2+\sqrt{M^4-J^2}\over\hbar}
\right]
\;\approx \;n\hbar \;;
\ee
\label{20}
which --- by way of Eq.(\ref{13}) --- can be reinterpreted as
the following quantum spectrum for the area:
\be
A_n \;-\;8\pi M^2\ln\left[{A_n\over 8\pi\hbar}\right]\; =\;
4\pi\hbar n\;.
\label{21}
\ee

The reader has most probably  noticed
the conspicuous presence of a logarithmic term in the spectrum.
Certainly, it is true that logarithmic
corrections to the horizon area  have had a long and storied
tradition in the literature \cite{page}.
But the key word here is ``correction''.
In Eq.(\ref{21}), the logarithmic term becomes the dominant one
for any black hole larger than (at most) a
few hundred Planck areas. Meanwhile, by invoking a {\it semi-classical}
quantization condition, we are --- by implication --- talking about
black holes with $A>>\hbar$. Hence, for the obligatory regime of validity,
the spectrum is far from evenly spaced (although still discrete).
To further complicate matters, $A>>\hbar$ ensures us that
the logarithm is positive, and so the
(typically dominant) logarithmic term is negative.
Meanwhile,  the left-hand side of the equation
is, in reality,  inherently positive!
[This follows directly from  the first and second laws of
thermodynamics; inasmuch as the integrand of
Eq.(\ref{18}) is essentially the (differential) change in entropy
of a rotating black hole, albeit with an unorthodox choice
for the temperature.] Finally, even if there could  be some
conceivable means of dismissing away the logarithmic
term, one would then  deduce a spacing
coefficient of $\epsilon=4\pi$ ---  disturbingly off by
a factor of two from
the previous Kerr result [{\it cf}, Eq.(\ref{17})].

So what exactly went wrong? The problem, as we see it, can be traced
to the evaluation of Eq.(\ref{19}), where the integration variables ($M$ and $J$)
have been placed on an equal footing. But such a democracy
of variables can not actually
be justified. To see this, let us first reconsider the
application of the Bohr--Sommerfeld quantization condition.
This immediately implies a semi-classical regime,
which requires $n$ to be a  very large number, as the
area  (in Planck units) must be as well.
Although not so immediately obvious, the domain of semi-classicality
indicates yet another constraint that must inevitably
be dealt with.
In short, the black hole must be far away from {\it extremality}.
(Keep in mind that the Kerr extremal limit is $J=M^2$. For
$J>M^2$, the black hole is replaced by a naked singularity,
in a  grievous  violation of cosmic censorship.)

To elucidate,
it has been demonstrated that, at least in the context
of Bekenstein-inspired  spectroscopy, a near-extremal
black hole (whether charged \cite{kun-2} and/or rotating
\cite{gour1}~/\cite{gour2})
 is a highly
quantum object. This is because, as clarified
in the cited articles, the
quantum number $n$  is actually  a measure
of the areal {\it deviation} from
extremality, rather than the area itself.
Meaning that a near-extremal black hole is necessarily
associated with a small quantum number, irrespective
of its overall size or mass.
Hence, for
the previous calculation to make sense, it is necessary
to restrict ourselves to small values of $J$; that is,
$J<<M^2$. It should be clear that no such restriction
has yet been
imposed.~\footnote{The reader may be concerned about
$n$ sometimes measuring  the area and other times, the
areal separation from extremality.  
However, in the true semi-classical
regime --- where $M^2>>J$ and so $A\approx 16\pi M^2$ --- this distinction
is inconsequential.}

A simple but  reasonable way to constrain the angular
momentum  is to treat
the integrand of Eq.(\ref{19}) as a perturbative expansion
in $J/M^2$. On this basis, one would be inclined to
 recast the expression
as follows:
\be
\int\left[4MdM \;-\;{J+{\cal O}(J^3)\over M^2}dJ \right]\;\approx \;
n\hbar\;.
\label{22}
\ee
The integration now leads to
\be
2M^2 -{1\over 2}{J^2\over M^2} +{\cal O}(J^4)
\;\approx \;n\hbar \;
\ee
\label{23}
or, equivalently,
\be
M^2+\sqrt{M^4-J^2}+{\cal O}(J^4) \;\approx \;n\hbar \;.
\ee
\label{24}
Comparing with Eq.(\ref{13}), we can now extrapolate
an area spectral  form of
\be
A_n +{\cal O}(J^4_n) \; =\;
8\pi\hbar n\;.
\label{25}
\ee
So that (for the semi-classical
domain of  both large $n$ and small $J\;$~\footnote{It should be noted that,
 because $J$ and its corresponding quantum number
are now being treated as parametrically small  quantities,
Eq.(\ref{25}) in no way contradicts any of  the
spectra documented in \cite{gour1,gour2}.}) the
Kerr spectral-spacing coefficient is reconfirmed
as taking on  the {\it universal} value of  $\epsilon=8\pi$. Moreover, the
once-dominant logarithmic term has  safely been  eradicated
and both sides of the equation
are now manifestly positive.

To summarize, we have found
that --- after fully accounting for  the
semi-classical  regime of
validity ---  the Kerr  area spectrum
is asymptotically identical for the two
distinct methods of interest. (These being Hod's \cite{hod-1} and
Kunstatter's \cite{kun-1}, along with the essential modifications
suggested by Maggiore \cite{mag}  and Vagenas \cite{vag}.)
In this way, we have also confirmed --- now  unambiguously ---
a universal form \cite{vag} for the Schwarzschild and
Kerr spectra.
 
\newpage

\section*{Acknowledgments}
Research is financially supported by the University of
Seoul. The author thanks Elias Vagenas for
useful discussions.


\end{document}